\documentclass{ws-procs9x6}

\begin{document}

\title{MAID Analysis Techniques\footnote{\uppercase{T}his work is supported by \uppercase{D}eutsche
\uppercase{F}orschungsgemeinschaft (\uppercase{SFB}443).}}

\author{Lothar~Tiator$^1$ and Sabit~Kamalov$^{1,2}$}

\address{
$^1$Institut f\"ur Kernphysik, Universit\"at Mainz, D-55099 Mainz, Germany\\
$^2$JINR Dubna, 141980 Moscow Region, Russia \\
E-mail: tiator@kph.uni-mainz.de, kamalov@kph.uni-mainz.de}

\maketitle

\abstracts{MAID is a unitary isobar model for a partial wave analysis of pion photo- and
electroproduction in the resonance region. It is fitted to the world data and can give
predictions for multipoles, amplitudes, cross sections and polarization observables in
the energy range from pion threshold up to $W=2$ GeV and photon virtualities $Q^2<5$
GeV$^2$. Using more recent experimental results from Mainz, Bates, Bonn and JLab for
$Q^2$ up to 4.0 GeV$^2$, the $Q^2$ dependence of the helicity couplings
 $A_{1/2}, A_{3/2}, S_{1/2}$
has been extracted for a series of four star resonances. We compare
single-$Q^2$ analyses with a superglobal fit in a new parametrization of Maid2005.
Besides the (pion) MAID, at Mainz we maintain a collection of online programs for partial
wave analysis of $\eta, \eta'$ and kaon photo- and electroproduction which are all based
on similar footings with field theoretical background and baryon excitations in
Breit-Wigner form.}

\section{Introduction}
\label{introduction}

In 1998 the first version of MAID (MAID98) was developed and implemented on the web for
an easy access for the whole community. MAID98 was based on a unitary isobar model
constructed with a limited set of the seven most important nucleon resonances for pion
photoproduction in Breit-Wigner form and a non-resonant background with Born terms and
t-channel vector meson exchange contributions.\cite{Maid} The model was unitarized for
each partial wave up to the $2\pi$ threshold, the region where Watson's theorem is
strictly valid. The model was later extended to eight resonances and the unitarization
procedure was modified in accordance with the dynamical model (MAID2000). But yet the
model was not fitted to the full world data base of pion photo- and electroproduction,
and some background parameters were adjusted to multipoles from the SAID partial wave
analysis. Results for finite $Q^2$ were mere predictions or extrapolations from
photoproduction, where only the magnetic form factor of the Delta excitation was obtained
from experimental results.

Since 2003 MAID has become a partial wave analysis program, where all parameters are
fitted to experimental observables as cross sections and polarization asymmetries from
pion photo- and electroproduction.

Besides the mostly developed Maid program for pion photo- and electroproduction on the
nucleon, at Mainz we have developed and collected a series of programs for photo- and
electroproduction of $\pi, \eta, K$ and $\eta'$, covering the whole pseudoscalar meson
nonet. The programs are available in the internet for online calculations under the URL
http://www.kph.uni-mainz.de/MAID. In detail we can report on the following status:

\begin{itemize}
\item {\bf MAID} is the main part of our project and describes the reaction $ e + N
\rightarrow e' + N + \pi$ in the kinematical range of $1.073$ GeV $< W < 2$ GeV and $Q^2
< 5$ GeV$^2$, see Refs.\cite{Maid,Maid05}. The current version is $Maid2005$.

\item {\bf DMT} (Dubna-Mainz-Taipei) is a dynamical model based on a previous work of the Taipei
group\cite{Yang85}. It also describes the reaction $ e + N \rightarrow e' + N + \pi$ in
the same kinematical range as Maid, see Ref. \cite{DMT,KY99}. The current version is
$DMT2001$.

\item {\bf KAON-MAID} is an isobar model developed by Mart and Bennhold. It describes the
reaction $ e + N \rightarrow e' + \{\Lambda,\Sigma\} + K$ and is applicable in the
kinematical range of $1.609$ GeV $< W < 2.2$ GeV  and $Q^2 < 2.2$ GeV$^2$, see
Ref.\cite{Lee99,Benn99}. The current version is $KaonMaid2000$.

\item {\bf ETA-MAID} is an isobar model for the reaction $ e + N \rightarrow e' + N +
\eta$. It exists in two versions, an older version ($EtaMaid2001$) that can give
predictions for photo- and electroproduction of $\eta$ from proton and neutron. It can be
used in the kinematical range of $1.486$ GeV $< W < 2.0$ GeV and $Q^2 < 5$ GeV$^2$, see
Ref.\cite{Chiang02}.

\noindent The more recent version ($EtaMaid2003$) incorporates in addition an option to
choose Regge tails for the t-channel vector meson exchange. It can only be applied to
photoproduction on the proton in the kinematical range of $1.486$ GeV $< W <$ 3.5 $GeV$
and $Q^2 < 5$ GeV$^2$, see Ref.\cite{Chiang03}.

\item {\bf ETA'-MAID} is an isobar model with t-channel $\omega,\rho$ Regge trajectories
for the reaction $\gamma + p \rightarrow p + \eta'$ in the kinematical range of $1.896$
GeV $< W < 3.5$ GeV and $Q^2 = 0$, see Ref.\cite{Chiang03}. The current version is
$EtaprimeMaid2003$.

\item {\bf DR-MAID} is a dispersion theoretical analysis of $ e + N
\rightarrow e' + N + \pi$ and is still in progress. It is based on fixed-t dispersion
relations and uses as an input the imaginary parts of the MAID amplitudes, see
Ref.\cite{DrMaid}. The current version $DrMaid2004$ is not yet available on the web.
\end{itemize}

\section{The dynamical approach to meson electroproduction}
\label{Photo- and electroproduction}

In the dynamical approach to pion photo- and electroproduction\cite{Yang85,KY99}, the
t-matrix is expressed as
\begin{eqnarray}
t_{\gamma\pi}(E)=v_{\gamma\pi}+v_{\gamma\pi}\,g_0(E)\,t_{\pi N}(E)\,, \label{eq:tgamapi}
\end{eqnarray}
where $v_{\gamma\pi}$ is the transition potential operator for $\gamma^*N \rightarrow \pi
N$, and $t_{\pi N}$ and $g_0$ denote the $\pi N$ t-matrix and free propagator,
respectively, with $E \equiv W$ the total energy in the CM frame. A multipole
decomposition of Eq. (\ref{eq:tgamapi}) gives the physical amplitude\cite{DMT}
\begin{eqnarray}
&&t_{\gamma\pi}^{(\alpha)}(q_E,k;E+i\epsilon)
=\exp{(i\delta^{(\alpha)})}\,\cos{\delta^{(\alpha)}} \times
[v_{\gamma\pi}^{(\alpha)}(q_E,k)\nonumber\\ && \qquad+P\int_0^{\infty} dq'
\frac{q'^2R_{\pi N}^{(\alpha)}(q_E,q';E)\,v_{\gamma\pi}^{(\alpha)}(q',k)}{E-E_{\pi
N}(q')}]\,, \label{eq:backgr}
\end{eqnarray}
where $\delta^{(\alpha)}$ and $R_{\pi N}^{(\alpha)}$ are the $\pi N$ scattering phase
shift and reaction matrix in channel $\alpha$, respectively; $q_E$ is the pion on-shell
momentum and $k=|{\bf k}|$ is the photon momentum. The multipole amplitude in Eq.
(\ref{eq:backgr}) manifestly satisfies the Watson theorem and shows that the $\gamma,\pi$
multipoles depend on the half-off-shell behavior of the $\pi N$ interaction.

In a resonant channel the transition potential $v_{\gamma\pi}$ consists of two terms
\begin{eqnarray}
v_{\gamma\pi}(E)=v_{\gamma\pi}^B +v_{\gamma\pi}^R(E),\label{eq:vgammapi33}
\end{eqnarray}
where $v_{\gamma\pi}^B$ is the background transition potential and $v_{\gamma\pi}^R(E)$
corresponds to the contribution of the bare resonance excitation. The resulting t-matrix
can be decomposed into two terms
\begin{eqnarray}
t_{\gamma\pi}(E)=t_{\gamma\pi}^B(E) + t_{\gamma\pi}^{R}(E)\,.\label{eq:tgammapi33}
\end{eqnarray}

The background  potential $v_{\gamma\pi}^{B,\alpha}(W,Q^2)$ is described by Born terms
obtained with an energy dependent mixing of pseudovector-pseudoscalar $\pi NN$ coupling
and t-channel vector meson exchanges. The mixing para\-meters and coupling constants were
determined from an analysis of nonresonant multipoles in the appropriate energy regions.
In the new version of MAID, the $S$, $P$, $D$ and $F$ waves of the background
contributions are unitarized in accordance with the K-matrix approximation,
\begin{equation}
 t_{\gamma\pi}^{B,\alpha}({\rm MAID})=
 \exp{(i\delta^{(\alpha)})}\,\cos{\delta^{(\alpha)}}
 v_{\gamma\pi}^{B,\alpha}(W,Q^2).
\label{eq:bg00}
\end{equation}

From Eqs. (\ref{eq:backgr}) and  (\ref{eq:bg00}), one finds that the difference between
the background terms of MAID and of the dynamical model is that off-shell rescattering
contributions (principal value integral) are not included in MAID, therefore, after
re-fitting the data, they are implicitly contained in the resonance part leading to
dressed resonances.

Following  Ref.\cite{Maid},  we assume a Breit-Wigner form for the resonance contribution
${\mathcal A}^{R}_{\alpha}(W,Q^2)$ to the total multipole amplitude,
\begin{equation}
{\mathcal A}_{\alpha}^R (W,Q^2)\,=\, \bar{\mathcal A}_{\alpha}^R (Q^2)\, \frac{f_{\gamma
R}(W)\Gamma_R\,M_R\,f_{\pi R}(W)}{M_R^2-W^2-iM_R\Gamma_R} \,e^{i\phi}, \label{eq:BW}
\end{equation}
where $f_{\pi R}$ is the usual Breit-Wigner factor describing the decay of a resonance
$R$ with total width $\Gamma_{R}(W)$ and physical mass $M_R$. The expressions for
$f_{\gamma R}, \, f_{\pi R}$ and $\Gamma_R$ are given in Ref.\cite{Maid}. The phase
$\phi(W)$ in Eq.(\ref{eq:BW}) is introduced to adjust the phase of the total multipole
to  equal  the corresponding $\pi N$  phase shift $\delta^{(\alpha)}$. While in the
original version of MAID only the 7 most important nucleon resonances were included with
mostly only transverse e.m. couplings, in our new version all four star resonances below
$W=2$~GeV are included. These are $P_{33}(1232)$, $P_{11}(1440)$, $D_{13}(1520)$,
$S_{11}(1535)$, $S_{31}(1620)$, $S_{11}(1650)$, $D_{15}(1675)$, $F_{15}(1680)$,
$D_{33}(1700)$, $P_{13}(1720)$, $F_{35}(1905)$, $P_{31}(1910)$ and $F_{37}(1950)$.

The resonance couplings $\bar{\mathcal A}_{\alpha}^R(Q^2)$ are independent of the total
energy and depend only on $Q^2$. They can be taken as constants in a single-Q$^2$
analysis, e.g. in photoproduction, where $Q^2=0$ but also at any fixed $Q^2$, where
enough data with W and $\theta$ variation is available. Alternatively they can also be
parametrized as functions of $Q^2$
in an ansatz like
\begin{equation}
\bar{\mathcal A}_{\alpha}(Q^2) = \bar{\mathcal A}_{\alpha}(0) (1+\beta_1 Q^2+\beta_2 Q^4
+ \cdots)\, e^{-\gamma Q^2}\,.
\end{equation}
With such an ansatz it is possible to determine the parameters $\bar{\mathcal
A}_{\alpha}(0)$ from a fit to the world database of photoproduction, while the parameters
$\beta_i$ and $\gamma$ can be obtained from a combined fitting of all electroproduction
data at different $Q^2$. The latter procedure we call the `superglobal fit'. In MAID the
photon couplings $\bar{\mathcal A}_{\alpha}$ are direct input parameters. They are
directly related to the helicity couplings $A_{1/2}, A_{3/2}$ and $S_{1/2}$ of nucleon
resonance excitation. For further details see Ref.\cite{Tiator03}.

\section{Data analysis}
\label{sec:data analysis}

The unitary isobar model MAID was used to analyze the world data of pion photo- and
electroproduction. In a first step we have fitted the background parameters of MAID and
the transverse normalization constants $\bar{\mathcal A}_{\alpha}^R (0)$ for the nucleon
resonance excitation. The latter ones give rise to the helicity couplings shown in Tables
1 and 2. Most of the couplings are in good agreement with PDG and the GW/SAID analysis.
It is very typical for such a global analysis, where about 15000 data points are fitted
to a small set of 10-20 parameters, that the fit errors appear unrealistically small.
Such errors only reflect the statistical uncertainty of the experimental errors, but the
model uncertainty can be ten times larger. Therefore we do not report these fit errors
which are very similar as in the GW02 fits. The only reliable error estimate can be
obtained by comparing different analyses like SAID, MAID and coupled channels analyses.

In Fig. 1 we give a comparison between MAID and SAID for three important multipoles,
$E_{0+}(S_{11}), M_{1-}(P_{11})$ and $E_{2-}(D_{13})$. For both analyses we show the
global (energy dependent) curves together with the local (single energy) fits, where only
data in energy bins of 10-20 MeV are fitted. Such a comparison demonstrates the
fluctuations due to a limited data base, especially in the case of the Roper multipole
$M_{1-}$. It also shows systematic differences between the MAID and SAID analyses in the
real parts of $E_{0+}$ and $E_{2-}$. Because of correlations between these amplitudes,
these differences cannot be resolved with our current data base. Because of isospin 1/2,
they can, however, lead to sizeable differences in the $\gamma,\pi^+$ channel, where the
data base is still quite limited.

\begin{table}[htbp]
\tbl{Proton helicity amplitudes at $Q^2=0$ of the major nucleon resonances. The results
from our own analyses with Maid2003 and the current Maid2005 version are compared to the
Particle Data Tables\protect\cite{PDG04} and the GW/SAID\protect\cite{SAID} analysis.
Numbers are given in units of $10^{-3}$ GeV$^{-1/2}$. \label{tab:1}}
{\begin{tabular}{|l|l|c|c|c|c|} \hline
  &  &  PDG & GW02 & MD03 & MD05 \\
\hline
$P_{33}(1232)$& $A_{1/2}$ & -135$\pm$6 & -129$\pm$1 & -140 & -137\\
              & $A_{3/2}$ & -255$\pm$8 & -243$\pm$1 & -265 & -260\\
\hline
$P_{11}(1440)$& $A_{1/2}$ & -65$\pm$4 & -67$\pm$4 & -77 & -61 \\
\hline
$D_{13}(1520)$& $A_{1/2}$ & -24 $\pm$9 & -24 $\pm$4 & -30  & -27\\
              & $A_{3/2}$ & 166 $\pm$5 & 135 $\pm$4 & 166  & 161\\
\hline
$S_{11}(1535)$& $A_{1/2}$ &  90$\pm$30 & 30$\pm$4 & 73 & 66  \\
\hline
$S_{31}(1620)$& $A_{1/2}$ &  27$\pm$11 &  -13$\pm$4 & 71   & 66  \\
\hline
$S_{11}(1650)$& $A_{1/2}$ &  53$\pm$16 & 74$\pm$4 & 32 & 33  \\
\hline
$D_{15}(1675)$& $A_{1/2}$ &  19 $\pm$8 &  33 $\pm$4 &  23  &  15\\
              & $A_{3/2}$ &  15 $\pm$9 &   9 $\pm$4 &  24  &  22\\
\hline
$F_{15}(1680)$& $A_{1/2}$ & -15 $\pm$6 & -13 $\pm$4 & -25  & -25\\
              & $A_{3/2}$ & 133 $\pm$12& 129 $\pm$4 & 134  & 134\\
\hline
$P_{13}(1720)$& $A_{1/2}$ &  18 $\pm$30&            &  55  &  73\\
              & $A_{3/2}$ & -19 $\pm$20&            & -32  & -11\\
\hline
\end{tabular}}
\end{table}

\begin{table}[htbp]
\tbl{Neutron helicity amplitudes at $Q^2=0$ of the major nucleon resonances. Notation as
in Table 1.\label{tab:2}}
{\begin{tabular}{|l|l|c|c|c|c|} \hline
  &  &  PDG & GW02 & MD03 & MD05 \\
\hline
$P_{11}(1440)$& $A_{1/2}$ & 40$\pm$10 & 47$\pm$5 & 52 & 54  \\
\hline
$D_{13}(1520)$& $A_{1/2}$ & -59 $\pm$9 & -67 $\pm$4 & -85  & -77\\
              & $A_{3/2}$ &-139 $\pm$11&-112 $\pm$3 &-148  &-154\\
\hline
$S_{11}(1535)$& $A_{1/2}$ & -46$\pm$27 & -16$\pm$5 & -42 & -51 \\
\hline
$S_{11}(1650)$& $A_{1/2}$ & -15$\pm$21 &-28$\pm$4 & 27 &  9  \\
\hline
$D_{15}(1675)$& $A_{1/2}$ & -43 $\pm$12& -50 $\pm$4 & -61  & -62\\
              & $A_{3/2}$ & -58 $\pm$13& -71 $\pm$5 & -74  & -84\\
\hline
$F_{15}(1680)$& $A_{1/2}$ &  29 $\pm$10&  29 $\pm$6 &  25  &  28\\
              & $A_{3/2}$ & -33 $\pm$9 & -58 $\pm$9 & -35  & -38\\
\hline
$P_{13}(1720)$& $A_{1/2}$ &   1 $\pm$15&            &  17  &  -3\\
              & $A_{3/2}$ & -29 $\pm$61&            & -75  & -31\\
\hline
\end{tabular}}
\end{table}

In a second step we have fitted recent differential cross section data on $p(e,e'p)\pi^0$
from Mainz\cite{Pos01}, Bates\cite{Mer01}, Bonn\cite{Ban02} and
JLab\cite{Lav01,Smith03,Fro99}. These data cover a $Q^2$ range from $0.1\cdots 4.0$
GeV$^2$ and an energy range $1.1$ GeV $< W < 2.0$ GeV. In a first attempt we have fitted
each data set at a constant $Q^2$ value separately. This is similar to a partial wave
analysis of pion photoproduction and only requires additional longitudinal couplings for
all the resonances. The $Q^2$ evolution of the background is described with nucleon Sachs
form factors in the case of the $s-$ and $u-$ channel nucleon pole terms. At the e.m.
vertices of the $\pi$ pole and seagull terms we apply the pion and axial form factors,
respectively, while a standard dipole form factor is used for the vector meson exchange.
Furthermore, as mentioned above, we have introduced a $Q^2$ evolution of the transition
form factors of the nucleon to $N^*$ and $\Delta$ resonances and have parameterized each
of the transverse $A_{1/2}$ and $A_{3/2}$ and longitudinal $S_{1/2}$ helicity couplings.
In a combined fit with all electroproduction data from the world data base of
GWU/SAID\cite{SAID} and the data of our single-$Q^2$ fit we obtained a $Q^2$ dependent
solution (superglobal fit).
\begin{figure}[htb]
\centerline{\epsfxsize=3.8in\epsfbox{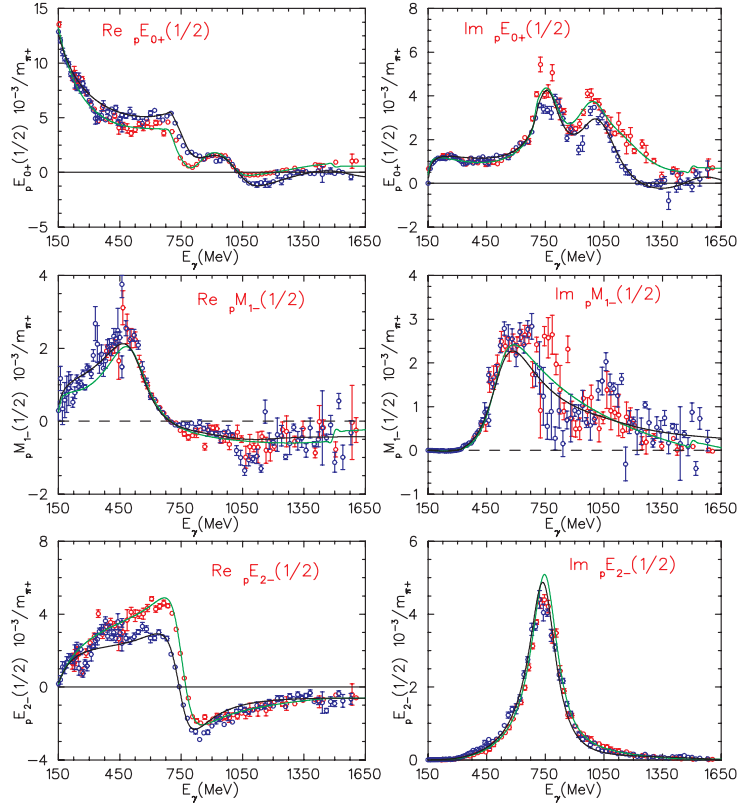}}
\caption{Comparison of MAID and SAID multipoles. The blue lines and points show the
Maid2005 global and single-energy solutions, respectively. The global and s.e. SAID
solutions are shown in green and red colours.}
\end{figure}
\label{fig:mult}

In Fig. 2 we show our results for the $\Delta(1232)$, the $D_{13}(1520)$ and the
$F_{15}(1680)$ resonances. Our superglobal fit agrees very well with our single-$Q^2$
fits, except in the case of the $\Delta$ resonance for the 2 lowest points of $S_{1/2}$
from our analysis of the Hall B data. Whether this is an indication for a different $Q^2$
dependence has still to be investigated. Generally, all our single-$Q^2$ points are shown
with statistical errors from $\chi^2$ minimization only. A much bigger error has to be
considered for model dependence.

\begin{figure}[htb]
\centerline{\epsfxsize=4.51in\epsfbox{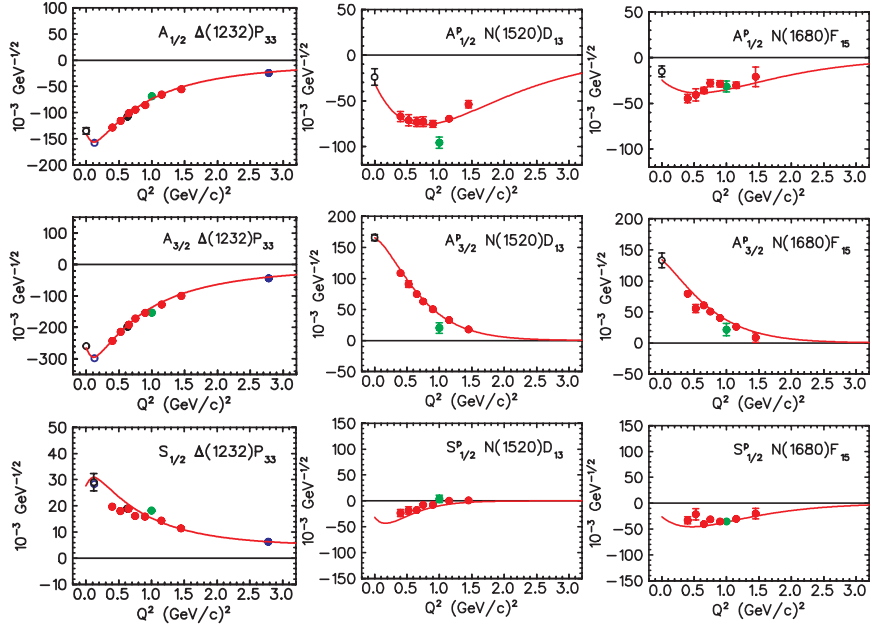}}
\caption{ The $Q^2$
dependence of the transverse ($A_{1/2},A_{3/2}$) and longitudinal ($S_{1/2}$) helicity
couplings for the $P_{33}(1232)$, $D_{13}(1520)$ and $F_{15}(1680)$ resonance excitation.
The solid curves show our superglobal fit. The data points at finite $Q^2$ are obtained
from our single-$Q^2$ analysis using the data from MAMI and Bates for $Q^2=0.1$GeV$^2$,
from ELSA for 0.6 GeV$^2$, JLAB(Hall A) for 1.0 GeV$^2$, JLab(Hall C) for 2.8 GeV$^2$ and
JLab(Hall B) for the remaining points. At the photon point ($Q^2=0$) we show our result
from Table 1 obtained from the world data base.}
\end{figure}
\label{fig:d13}

In Fig. 3 we show our results for the helicity amplitudes of the Roper resonance
$P_{11}(1440)$ and the $S_{11}(1535)$ in the region of $Q^2 < 1$ GeV$^2$. In addition to
our own singe-$Q^2$ analysis we also compare to the analysis of Aznauryan and
Burkert\cite{Aznau05} who used both an isobar model similar to Maid and an analysis based
on fixed-t dispersion relation. In general we get a good agreement with the results of
Ref.\cite{Aznau05}. Only for the longitudinal excitation of the $S_{11}$ resonance one
may observe a different tendency of the $Q^2$ dependence, however, in this case the
statistical fluctuations of our analysis is quite large.

\begin{figure*}[ht]
\centerline{\epsfxsize=4.0in\epsfbox{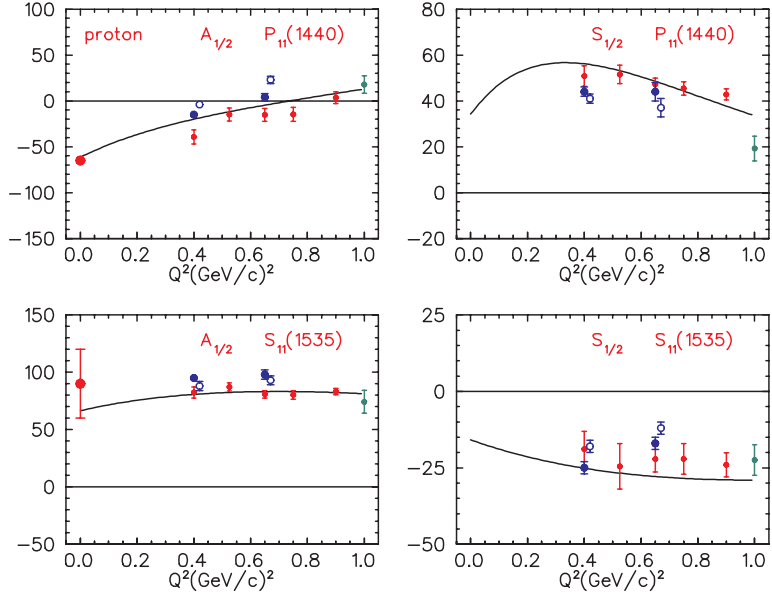}}
\caption{The $Q^2$ dependence of the transverse and longitudinal helicity amplitudes
for the $P_{11}(1440)$ and the $S_{11}(1535)$ resonance excitation of the proton. The
solid lines are the superglobal Maid2005 solutions. The solid red (gray) points are our
single-$Q^2$ fits to the exp. data from CLAS/JLab\protect\cite{Smith03}, the solid and
open blue circles show the isobar and dispersion analysis of Aznauryan
\protect\cite{Aznau05} using a similar data set.}
\end{figure*}
\label{fig:p11s11} Furthermore we also have some empirical results for the partial waves
that are not shown here, but most of them come out of the fit with rather large errors
bars in the single-$Q^2$ analysis. This gives us less confidence also for our superglobal
fit. The reason for it is mainly that we have fewer data points to analyze at higher
energies.

\section{Conclusions}
\label{sec:concl}

Using the world data base of pion photo- and electroproduction and recent data from
Mainz, Bonn, Bates and JLab we have made a first attempt to extract all longitudinal and
transverse helicity amplitudes of nucleon resonance excitation for four star resonances
below $W=2$~GeV. For this purpose we have extended our unitary isobar model MAID and have
parameterized the $Q^2$ dependence of the transition amplitudes. Comparisons between
single-$Q^2$ fits and a $Q^2$ dependent superglobal fit give us confidence in the
determination of the helicity couplings of the $P_{33}(1232), P_{11}(1440), S_{11}(1535),
D_{13}(1520)$ and the $F_{15}(1680)$ resonances, even though the model uncertainty of
these amplitudes can be as large as 50\% for the longitudinal amplitudes of the $D_{13}$
and $F_{15}$.

For other resonances the situation is more uncertain. However, this only reflects the
fact that precise data in a large kinematical range are absolutely necessary. In some
cases double polarization experiments are very helpful as has already been shown in pion
photoproduction. Furthermore, without charged pion electroproduction, some ambiguities
between partial waves that differ only in isospin as $S_{11}$ and $S_{31}$ cannot be
resolved without additional assumptions. While all electroproduction results discussed
here are only for the proton target, we have also started an analysis for the neutron,
where much less data are available from the world data base and no new data has been
analyzed in recent years. Since we can very well rely on isospin symmetry, only the
electromagnetic couplings of the neutron resonances with isospin $1/2$ have to be
determined. We have obtained a global solution for the neutron which is implemented in
MAID2005. However, for most resonances this is still highly uncertain. So it will be a
challenge for the experiment to investigate also the neutron resonances in the near
future.

\section{Acknowledgements}
We wish to thank Cole Smith for having access on recent experimental data. This work was
supported in part by the Deutsche Forschungsgemeinschaft (SFB443).


\begin{thebibliography}{}

\bibitem{Maid}D. Drechsel, O. Hanstein, S.S. Kamalov and L. Tiator,\\
Nucl. Phys. {\bf A645} (1999) 145;
  http://www.kph.uni-mainz.de/MAID/.
\bibitem{Maid05}D. Drechsel, S.S. Kamalov and L. Tiator, Maid2005, to be published.
\bibitem{Yang85}S.N. Yang, J. Phys. G {\bf 11} (1985) L205.
\bibitem{DMT}S. Kamalov, S.N. Yang, D. Drechsel, O. Hanstein, L. Tiator,\\
       Phys. Rev. C {\bf 64} (2001) 032201.
\bibitem{Lee99} F.X. Lee, T. Mart, C. Bennhold, H. Haberzettl, L.E. Wright,\\
Nucl. Phys. {\bf A695} (2001) 237.
\bibitem{Benn99} C. Bennhold, H. Haberzettl and T. Mart, Proc. of 2nd ICTP Int. Conf.
 on Perspectives in Hadronic Physics, Trieste 1999, p. 328, nucl-th/9909022.
\bibitem{Chiang02} W.-T. Chiang, S.N. Yang, L. Tiator and D. Drechsel,\\
Nucl. Phys. {\bf A700} (2002) 429.
\bibitem{Chiang03} W.-T. Chiang, S.N. Yang, L. Tiator, M. Vanderhaeghen and D. Drechsel, Phys. Rev.
  C {\bf 68} (2003) 045202.
\bibitem{DrMaid} S.S. Kamalov, L. Tiator, D. Drechsel, R.A. Arndt, C. Bennhold, I.I.
  Strakovsky and R.L. Workman, Phys. Rev. C {\bf 66} (2000) 065206.
\bibitem{KY99} S.S. Kamalov and S.N. Yang, Phys. Rev. Lett. {\bf 83} (1999) 4494.
\bibitem{Tiator03} L. Tiator, D. Drechsel, S. Kamalov, M.M. Giannini, E. Santopinto,
   and A. Vassallo, EPJ A {\bf 19} (2004) 55.
\bibitem{PDG04} S. Eidelman {\it et al.} (Particle Data Group), Phys. Lett. B {\bf 592} (2004) 1.
\bibitem{SAID} R.A. Arndt, W.J. Briscoe, I.I.Strakovsky, and R.L. Workman,\\
Phys. Rev. C {\bf 66} (2002) 055213; http://gwdac.phys.gwu.edu/.
\bibitem{Pos01}Th. Pospischil {\it et al.}, Phys. Rev. Lett. {\bf 86}
  (2001) 2959.
\bibitem{Mer01}C. Mertz {\it et al.}, Phys. Rev. Lett. {\bf 86}
  (2001) 2963.
\bibitem{Ban02}T. Bantes, PhD thesis Bonn 2003, BONN-IR-2003-08. 
\bibitem{Lav01} G. Laveissiere {\it et al.}, Phys. Rev. C {\bf 69} (2004) 045203.
\bibitem{Smith03}K. Joo {\it et al.}, Phys. Rev. Lett. {\bf 88}
  (2002) 122001-1. 
\bibitem{Fro99}V.V. Frolov {\it et al.}, Phys. Rev. Lett. {\bf 82} (1999) 45.
\bibitem{Aznau05} I. Aznauryan, V. Burkert, H. Egiyan, K. Joo, R. Minehart and L.C. Smith,
  Phys. Rev. C {\bf 71} (2005) 015201.


\end{thebibliography}
\end{document}